\documentclass{ws-ijmpe}

\begin{document}

\markboth{R. \'Alvarez-Rodr\'{\i}guez, A.S. Jensen,
D.V. Fedorov. H.O.U. Fynbo and E. Garrido}{$\alpha$ particle momentum
distributions from $^{12}$C decaying resonances}

\catchline{}{}{}{}{}
%%%%%%%%%%%%%%%%%%%%%%%%%%%%%%%%%%%%%%%%%%%%%%%%%%%%%%%%%%%%%%%%%%%%

\title{$\alpha$ PARTICLE MOMENTUM DISTRIBUTIONS FROM $^{12}$C DECAYING
RESONANCES}

\author{\footnotesize R. \'ALVAREZ-RODR\'IGUEZ, A.S. JENSEN,
D.V. FEDOROV, H.O.U. FYNBO}

\address{Department of Physics and Astronomy, University of Aarhus \\Ny
Munkegade Bygning 1520, DK-8000 Aarhus C, Denmark, \\ 
raquel@phys.au.dk}

\author{E. GARRIDO}

\address{Instituto de Estructura de la Materia, Consejo Superior de
Investigaciones Cient\'{\i}ficas\\ Serrano 123, E-28006 Madrid, Spain
}

\maketitle

\begin{history}
\received{(received date)}
\revised{(revised date)}
%\accepted{(Day Month Year)}
%\comby{(xxxxxxxxxx)}
\end{history}

\begin{abstract}
The computed $\alpha$ particle momentum distributions from the decay
of low-lying $^{12}$C resonances are shown. The wave function of the
decaying fragments is computed by means of the complex scaled
hyperspherical adiabatic expansion method.  The large-distance part of
the wave functions is crucial and has to be accurately calculated. We
discuss energy distributions, angular distributions and Dalitz plots
for the $4^+$, $1^+$ and $4^-$ states of $^{12}$C.
\end{abstract}

\section{Introduction}

The $^{12}$C resonances below the proton separation threshold have
been intensively studied over many years, motivated partly by its
astrophysical importance. There are still many unanswered questions,
e.g., what are the energies, angular momenta, structure, and decay
properties of the resonances. Open questions still remain on the $0^+$
and $2^+$ resonances. The existence of the latter was conjectured by
Morinaga in the fifties as a member of the rotational band with the
$0^+$ resonance at 7.65~MeV as band-head\cite{mor56}. Several
experiments recently provided new results but unfortunately yet for
the position and width of the first $2^+$ resonance, no agreement has
been reached.

Three-body decay is important for studying different decay mechanisms,
i.e. direct versus sequential. Direct decay takes place when all three
particles leave simultaneously their interaction regions, while
sequential decay proceeds via an intermediate 2-body state. The
intermediate path of the decay process is not an observable, therefore
the information has to be extracted from the distribution of the
fragments after the decay.

We investigate in this contribution the decay of low-lying continuum
states into three particle final states for the case of $^{12}$C,
assuming that the decay mechanism is independent of how the initial
state was formed. We describe the decay in analogy with
$\alpha$-decay, assuming that the three fragments are formed before
entering the barrier at sufficiently small distances to allow the
three-body treatment. Outside the range of the strong interaction,
only the Coulomb and centrifugal barriers remain, since we have
assumed that the small distance many-body dynamics is unimportant for
the process. We show the momentum distributions of the fragments after
the decay of the resonances by means of energy distributions, Dalitz
plots and angular distributions. A comparison with oncoming
experimental data is straightforward.

\section{Faddeev equations and complex scaling}

The decay of a $^{12}$C resonance into three $\alpha$ particles is a
pure three-body problem of nuclear physics. Therefore we describe
$^{12}$C as a 3$\alpha$-cluster system. We use Faddeev equations and
solve them in coordinate space using the adiabatic hyperspherical
expansion method\cite{nie01,gar05b,fed03}. The hyperspherical
coordinates consist in the so-called hyperradius $\rho$, defined as
\begin{equation}
\rho^2 = 4 \sum_{i=1}^3 (\vec r_i - \vec R)^2 \:,
\end{equation}
where $\vec r_i$ is the coordinate of the $i$-th $\alpha$ particle and
$\vec R$ is the coordinate of the centre of mass, and five generalised
angles. The angular functions are chosen for each $\rho$ as the
eigenfunctions of the angular part of the Faddeev equations. We must
first determine the interaction $V_i$ reproducing the low-energy
two-body scattering properties. In this case, we have chosen an
Ali-Bodmer potential\cite{ali66} slightly modified in order to
reproduce the s-wave resonance of $^8$Be. The energy of the resonance
is corrected by including a diagonal three-body interaction $V_{3b} =
S \exp (-\rho^2 / b^2)$. This three-body potential should effectively
mock up the transition between the N- and three-body degrees of
freedom, while the structure of the resonance is maintained.

The momentum distribution of the decay fragments is determined by the
Fourier transform of the coordinate-space wave function. The
hyperspherical harmonics transform into themselves after Fourier
transformation. It has been shown\cite{fed04} that the angular
amplitude of the momentum-space wave function of the resonance is
directly proportional to the coordinate-space one for a large value of
$\rho$. Numerically converged results in the appropriate region of
$\rho$-values are then needed in order to have a reliable computation.
The probability distribution is obtained after integration over the
four hyperangles describing the directions of the momenta,
\begin{equation}
P(k_y^2) \propto P(\cos^2 \alpha) \propto (\sin 2\alpha)\int d\Omega_x
\: d\Omega_y \: |\Psi (\rho,\alpha,\Omega_x,\Omega_y)|^2 \:.
\end{equation}
The asymptotic behaviour is reached for hyperradii larger than about
60~fm. There is a small variation of the distribution from 70 to
100~fm, that shows the stability and convergence of the
computation. We have chosen 80~fm as the value of $\rho$ where the
energy distributions should be computed.  We have performed a Monte
Carlo integration over the phase space to get the probability
distributions.

\section{Energy distributions and Dalitz plots}

\begin{figure}[th]
\vspace*{-28pt}
\centerline{\psfig{file=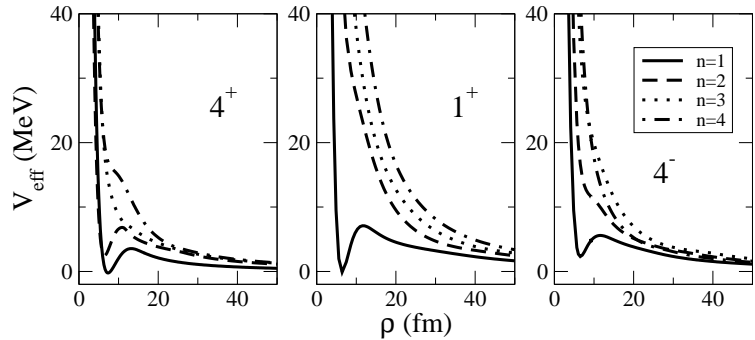,width=9cm,angle=270}}
\vspace*{-90pt}
\caption{The real parts of the four lowest adiabatic effective
potentials (labeled by $n$), including the three-body potentials, 
for the $4^+$ (left),
$1^+$ (centre) and $4^-$ (right) resonances of $^{12}$C. }
\label{figpot}
\end{figure}

We find fourteen $^{12}$C resonances below the proton separation
threshold for most angular momenta and both parities. Small and
intermediate distances are important for energies, widths and partial
wave decomposition. Their structures have been previously described in
detail\cite{alv07a}. Fig.~\ref{figpot} shows the adiabatic effective
potentials for the resonances we are considering here, i.e. $4^+$,
$1^+$ and $4^-$. At relatively small distances the potentials have
minima allocating the resonances. The barrier at intermediate
distances determines the width of the resonance.

The energy distribution of the fragments after the decay is the only
experimental information one can get, that allows us to study the
decay path. Dalitz plots contain more information than single $\alpha$
energy distributions and constitute an easy way to see how the three
particles share the energy after the decay of the resonance. The
angular distribution reflects the preferred direction followed by one
of the $\alpha$ particles with respect to the direction between the
other two. It must also reflect the behaviour of the angular momentum
and can be used to assign spin and parity of a measured
state\cite{alv07d}.

The {\it natural-parity states} of $^{12}$C, i.e.  $0^+$, $1^-$,
$2^+$, $3^-$ and $4^+$, can breakup in a sequential decay via
$^8$Be($0^+$). By exploiting the fact that precisely one of the
adiabatic potentials asymptotically must describe the two-body
resonance and the third particle far away, we are able to estimate the
amount of sequential decay by looking at the complex scaled radial
wave functions. 

\begin{figure}[th]
\centerline{\psfig{file=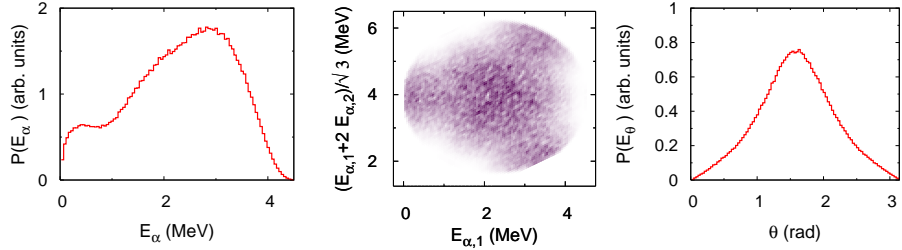,width=12.5cm}}
\vspace*{8pt}
\caption{$\alpha$ particle energy distribution (left), Dalitz plot
(centre) and angular distribution of the directions between two
particles and their centre of mass and the third particle (right) for
the $4^+$-resonance of $^{12}$C at 6.83~MeV above the 3$\alpha$
threshold at 7.275~MeV. The contribution from the decay through
$^8$Be$(0^+)$ has been removed. }
\label{fig4plus}
\end{figure}

For the $4^+$ state at an excitation energy of 14.1~MeV (or 6.83~MeV
above the 3$\alpha$ threshold) we estimate that 20\% decays via
the $^8$Be ground state. In fig.~\ref{fig4plus} we show the energy
distribution for an $\alpha$ particle, the Dalitz plot and the angular
distribution after removal of the sequential decay though the $^8$Be
ground state. The distribution of the kinetic energy of the particles
is rather diffuse, and the Dalitz plot is smeared out. This is in
contrast with the sequential decay distribution, where one of the
particles stays at high energy and the other two at low energy. The
angular distribution exhibits one smooth peak around $\pi$/2.

For the {\it unnatural-parity states} of $^{12}$C, i.e.  $1^+$, $2^-$,
and $4^-$, angular momentum and parity conservation forbid the decay
via $^8$Be($0^+$).

\begin{figure}[th]
\centerline{\psfig{file=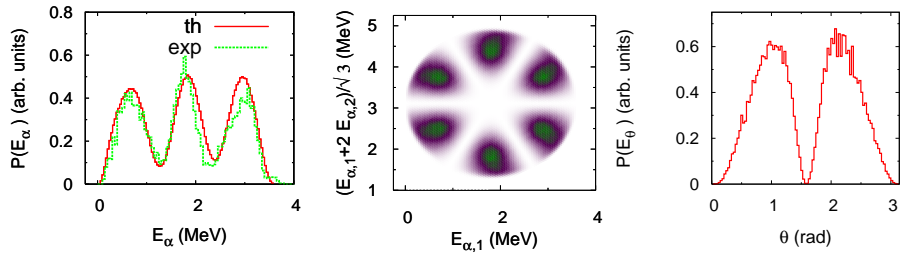,width=12.5cm}}
\vspace*{8pt}
\caption{$\alpha$ particle energy distribution (left), Dalitz plot
(centre) and angular distribution of the directions between two
particles and their centre of mass and the third particle (right) for
the $1^+$-resonance of $^{12}$C at 4.52~MeV above the 3$\alpha$
threshold at 7.275~MeV. }
\label{fig1plus}
\end{figure}

Fig.~\ref{fig1plus} shows the individual $\alpha$ particle energy
distribution, the Dalitz plot and the angular distribution for the
$1^+$ resonance of $^{12}$C. The $1^+$ state at an excitation energy
of 12.7~MeV (or 5.42~MeV above the 3$\alpha$ threshold) is referred to
as a shell-model state, which means that it has no cluster
structure. But at large distances, after the decay takes place, we are
dealing with a three-body problem. A 3$\alpha$ cluster description
seems then to be the most natural treatment for this system. For the
single $\alpha$ energy distribution the agreement with the
experiment\cite{diget} is almost perfect\cite{alv07b}. The theoretical
Dalitz plot can also be compared with the experimental
one\cite{hans}. It is clear that our computation reproduces the
pattern obtained from the experimental data. In the angular
distribution we obtain a minimum at $\pi/2$, which reflects the
intrinsic angular momenta used to construct the wave function.

\begin{figure}[th]
\centerline{\psfig{file=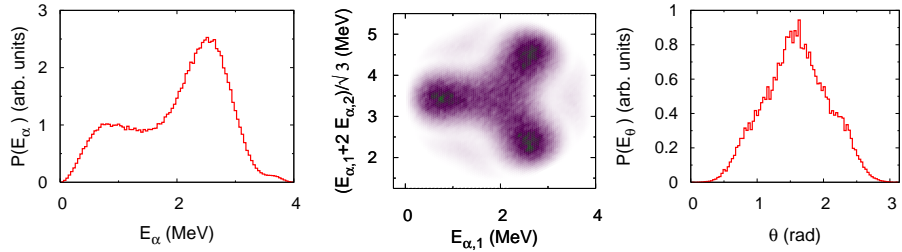,width=12.5cm}}
\vspace*{8pt}
\caption{$\alpha$ particle energy distribution (left), Dalitz plot
(centre) and angular distribution of the directions between two
particles and their centre of mass and the third particle (right) for
the $4^-$-resonance of $^{12}$C at 5.98~MeV above the 3$\alpha$
threshold at 7.275~MeV. }
\label{fig4minus}
\end{figure}

In our computation we obtain one $2^-$ and one $4^-$ state.  We
suggest the latter should correspond to the state at 13.35~MeV of
excitation energy (or 6.08~MeV above the 3$\alpha$
threshold)\cite{alv07c} for which a preliminary spin-parity of $2^-$
is assigned by Azjenberg-Selove\cite{azj}. A new spin-parity
assignment regarding the $2^-$ and $4^-$ states has been also recently
suggested by Freer et al.\cite{fre07}.

Fig.~\ref{fig4minus} shows the $\alpha$ energy distribution after the
decay of the $4^-$ resonance of $^{12}$C. It consists of a low peak at
small energies and a higher peak at larger energies. Oddly we get a
very similar energy distribution for the $2^-$ resonance at
11.8~MeV. However, both the Dalitz plots and the angular distributions
differ more from each other\cite{alv07d}.

We have compared our results with some preliminary data from the
reaction $^{11}$B($^3$He,d$\alpha\alpha\alpha$) studied at CMAM
(Madrid) by M.~Alcorta and collaborators in 2008\cite{martin}. Even
though the analysis of the data is not yet finished, we find a very
good agreement between the measured and the theoretical energy
distributions, angular distributions and Dalitz plots.

\section{Summary and conclusions}

We predict the $\alpha$-particle momentum distributions of the
decaying low-lying $^{12}$C many-body resonances. We use a pure
three-$\alpha$ cluster model to describe all states at all distances.
We mock up the small-distance many-body properties by a structureless
three-body interaction adjusted to reproduce the resonance energy.

The natural-parity state $4^+$ and the unnatural-parity states $1^+$
and $4^-$ are used as examples. We show $\alpha$ particle energy
distributions, Dalitz plots and angular distributions. Preliminary
experimental data seem to agree very nicely with the theoretical
predictions.

\section*{Acknowledgements}

R.A.R. acknowledges support by a post-doctoral fellowship from
Ministerio de E\-du\-ca\-ci\'on y Ciencia (Spain). The authors would like to
thank M. Alcorta and O. Tengblad for providing their preliminary
experimental data.


\begin{thebibliography}{0}
%\bibitem{1} J. Callaway, {\it Phys. Rev. B} {\bf 35} (1987) 8723.


\bibitem{mor56} H. Morinaga, 
{\it Phys. Rev.} {\bf 101} (1956) 254.

\bibitem{nie01} E. Nielsen, D.V. Fedorov, A.S. Jensen, and E. Garrido,
{\it Phys. Rep.} {\bf 347} (2001) 373.

\bibitem{gar05b} E. Garrido, D.V. Fedorov, A.S. Jensen and H.O.U. Fynbo,
{\it Nucl. Phys. A}  {\bf 766} (2005) 74.

\bibitem{fed03}  D.V. Fedorov, E. Garrido, and A.S. Jensen,
{\it Few-body systems} {\bf 33} (2003) 153.

\bibitem{ali66}  S. Ali and A.R. Bodmer,
{\it Nucl. Phys.} {\bf 80} (1966) 99.

\bibitem{fed04} D.V. Fedorov, H.O.U. Fynbo, E. Garrido and A.S. Jensen,
{\it Few-body systems} {\bf 34} (2004) 33.

\bibitem{alv07a} R. \'Alvarez-Rodr\'{\i}guez, E. Garrido, A.S. Jensen, 
D.V. Fedorov and H.O.U. Fynbo,
{\it Eur. Phys. J. A} {\bf 31} (2007) 303.

\bibitem{alv07d} R. \'Alvarez-Rodr\'{\i}guez, A.S. Jensen, E. Garrido,
D.V. Fedorov and H.O.U. Fynbo, 
{\it Phys. Rev. C} {\bf 77} (2008) 064305.

\bibitem{diget} C.Aa. Diget, Phd Thesis, University of Aarhus (2006).

\bibitem{alv07b} R. \'Alvarez-Rodr\'{\i}guez, A.S. Jensen, D.V. Fedorov,
H.O.U. Fynbo and E. Garrido,
{\it Phys. Rev. Lett.} {\bf 99} (2007) 072503.

\bibitem{hans} H.O.U. Fynbo, Y. Prezado, U.C. Bergmann, M.J.G. Borge,
P. Dendooven, W.X. Huang, J. Huikari, H. Jeppesen, P. Jones,
B. Jonson, M. Meister, G. Nyman, K. Riisager, O. Tengblad,
I.S. Vogelius, Y. Wang, L. Weissman, K. Wilhelmsen Rolander, and
J. \"Ayst\"o,
{\it Phys. Rev. Lett.} {\bf 91} (2003) 082502.

\bibitem{alv07c} R. \'Alvarez-Rodr\'{\i}guez, E. Garrido, A.S. Jensen, 
D.V. Fedorov and H.O.U. Fynbo,
{\it J. Phys. G: Nucl. Part. Phys.} {\bf 35} (2008) 014010.

\bibitem{azj} F. Azjenberg-Selove, 
{\it Nucl. Phys. A} {\bf 506} (1990) 1.

\bibitem{fre07} M. Freer, I. Boztosun, C.A. Bremner, S.P.G. Chappell, 
R.L. Cowin, G.K. Dillon, B.R. Fulton, B.J. Greenhalgh, T. Munoz-Britton, 
M.P. Nicoli, W.D.M. Rae, S.M. Singer, N. Sparks, D.L. Watson and D.C. Weisser,
{\it Phys. Rev. C} {\bf 76} (2007) 034320.

\bibitem{martin} M. Alcorta, private communication; M. Alcorta et al. 
To be published.



\end{thebibliography}
\end{document}